# Magnetic frustration in a van der Waals metal CeSiI

Ryutaro Okuma,[1,4] Clemens Ritter,[2] Gøran J. Nilsen,[3] Yoshinori Okada[1]

[1]*Okinawa Institute of Science and Technology Graduate University, Onna-son, Okinawa, 904-0495, Japan*
[2]*Institut Laue-Langevin, 71 avenue des Martyrs, 38042 Grenoble, France*
[3]*ISIS Neutron and Muon Source, Science and Technology Facilities Council, Didcot OX11 0QX, United Kingdom*
[4]*Present address: Clarendon Laboratory, University of Oxford, Oxford, OX1 3PU, UK*

The realization of magnetic frustration in a metallic van der Waals (vdW) coupled material has been sought as a promising platform to explore novel phenomena both in bulk matter and in exfoliated devices. However, a suitable material platform has been lacking so far. Here, we demonstrate that CeSiI hosts itinerant electrons coexisting with exotic magnetism. In CeSiI, the magnetic cerium atoms form a triangular bilayer structure sandwiched by van der Waals stacked iodine layers. From resistivity and magnetometry measurements, we confirm the coexistence of itinerant electrons with magnetism with dominant antiferromagnetic exchange between the strongly Ising-like Ce moments below 7 K. Neutron diffraction directly confirms magnetic order with an incommensurate propagation vector $k \sim (0.28, 0, 0.19)$ at 1.6 K, which points to the importance of further neighbor magnetic interactions in this system. The presence of a two-step magnetic-field-induced phase transition along *c* axis further suggests magnetic frustration in the ground state. Our findings provide a novel material platform hosting a coexistence of itinerant electron and frustrated magnetism in a vdW system, where exotic phenomena arising from rich interplay between spin, charge and lattice in low dimension can be explored.

The exploration of novel quantum phenomena in van der Waals (vdW) coupled materials is a promising and rapidly growing field motivated by both fundamental physics interest and, thanks to recent advances in exfoliation technologies, device applications[1-4]. An essential component to accelerate the development of the field is to increase the number of materials that host states whose properties are described by multiple coupled degrees of freedom, such as charge, spin, orbital, and lattice[5]. Among these, a particularly interesting playground for discovering such emergent phenomena has been identified in systems which manifest an interplay between itinerant electrons and magnetism, and intensive studies have been pursued and reported[6-17]. However, a promising area of this playground which simultaneously remains relatively unexplored is that of vdW materials, where itinerant electrons coexist with low-dimensional and potentially frustrated magnetism. Magnetic frustration can result in non-collinear spin textures and field-induced transitions to exotic quantum phases[18,19]. This includes emergent topologically non-trivial spin textures which lead to exotic couplings between the magnetism and the motion of itinerant electrons[20,21]. Furthermore, non-collinear spin textures can couple to lattice degrees of freedom to produce phenomena like multiferroicity[22].

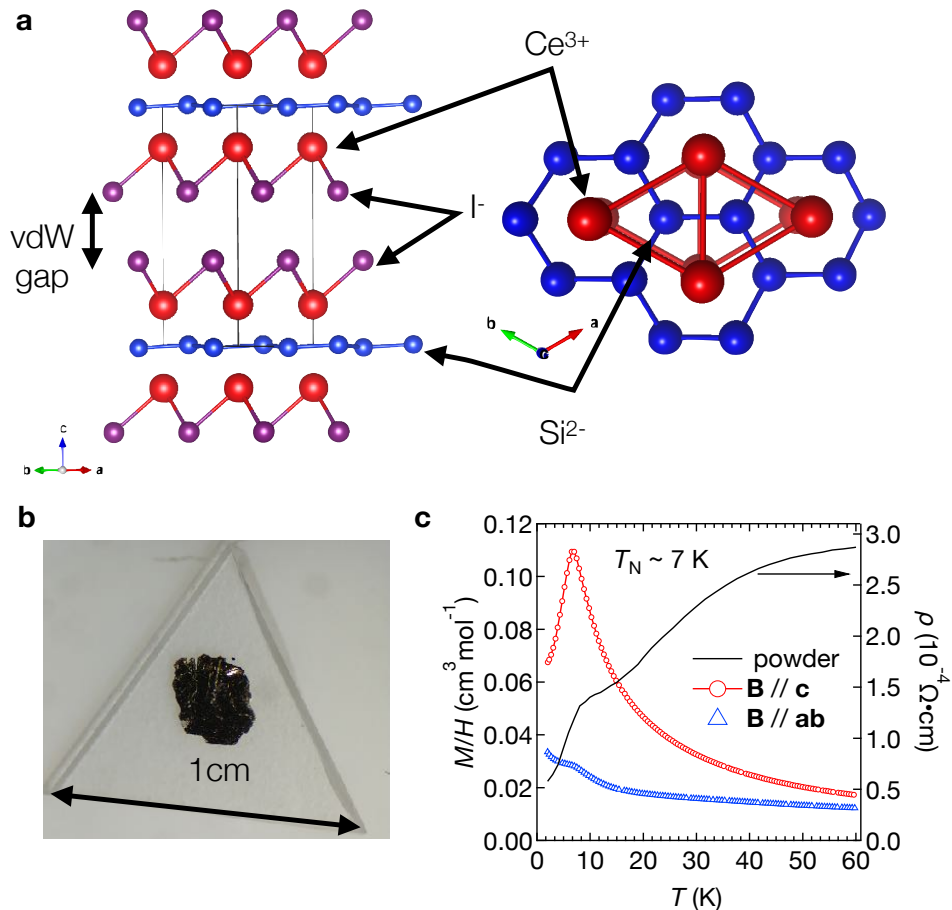

FIG.1. | **a,** Crystal structure of CeSiI. Red, blue, and purple spheres indicate Ce, silicon, and iodine atoms, respectively. Red, green, and blue arrows indicate the *a*, *b*, and *c* axes, respectively. **b,** Optical microscope image of a single crystal of CeSiI. The image is taken in a globe box filled with pure Ar gas. **c,** Temperature dependence of magnetic susceptibilities from a single crystal of CeSiI and zero-field resistivity of a polycrystalline sample of CeSiI. The magnetization measurements were carried out upon cooling in a magnetic field of 1 T applied along the *a* and *c* axes.

Part of the reason why magnetic metallic vdW materials are so scarce is that most vdW materials studied so far have been chalcogenides, which tend to be nonmagnetic. Another important class of materials with vdW structures is the halides. The large ionic radius and small number of chemical bonds of halides leads to low dimensional vdW bonded structures[23]. While typical transition metal halides[24-27] are low-dimensional Mott-insulators, this is not the case for reduced rare-earth halides $LnX_2$. In rare-earth elements with a strong tendency to stabilize the trivalent state, an electron produced by reduction occupies the outer 5*d* orbital of lanthanide rather than inner 4*f* orbital and metallicity results[28-30]. Moreover, the finding of giant magnetoresistance and ferromagnetic order above room temperature in GdI$_2$ motivates us to expect that 4*f* magnetism strongly couples to conduction electrons in rare earth halide materials[31,32].

The ternary system *LnA*I (*Ln* = La, Ce, Pr, Gd. *A* = Al, Si, Ga, Ge) is a class of materials closely related to the reduced rare earth halides[33-35]. CeSiI appears as a cleavable material in 2DMatpedia[36], and its crystal structure is depicted in **Fig.1a**. Ce atoms form a triangular bilayer in the *ab* plane sandwiching a honeycomb net of Si. These blocks are terminated by a vdW-coupled layer of iodine. While its synthesis and crystal structure have been reported previously, its physical properties, especially the effect of geometrical frustration and the influence of the Fermi surface on magnetism, have never been investigated. In order to demonstrate CeSiI as a promising host of itinerant electrons with frustrated magnetism, we have synthesized crystalline CeSiI by a high-temperature-solid-state reaction[33] in a bulk form (**Fig.1b**) and performed resistivity, magnetometry, heat capacity, and neutron diffraction measurements.

The coexistence of metallicity and antiferromagnetic interactions in a CeSiI is confirmed. **Figure 1c** shows the temperature-dependence of the susceptibility $\chi(T)$ along the *a* and *c* axes (left axis), together with the temperature dependence of the resistivity (right axis). $\chi(T)$ exhibits strongly anisotropic behavior below 100 K due to the trigonal crystalline electric field splitting of the $J = 5/2$ manifold of the $Ce^{3+}$ 4*f* electrons. The lowest Kramers doublet dominates the magnetism below ~10 K because the other $J = 5/2$ levels are typically located above $10^2$~$10^3$ K in the trigonal crystalline electric field[37]. Below 7 K, the susceptibility component along the *c* axis drops, while that along *a* axis slightly increases, which indicates long-range antiferromagnetic ordering at $T_N$ ~ 7 K. The presence of a magnetic phase transition is also confirmed by the drop of resistivity below 7 K due to suppression of electron-spin scattering.

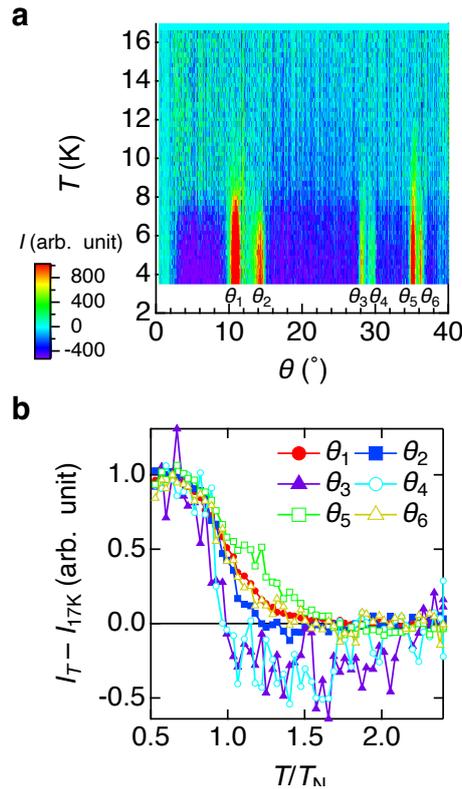

**FIG.2.** | Neutron diffraction study of CeSiI. (a) Temperature dependence of magnetic diffraction in the temperature range from 16 to 3.8 K. Nuclear reflections were removed from the data by subtracting data obtained at 17 K. The positions of the six observed magnetic Bragg peaks are labeled $\theta_i$ (i = 1, ..., 6). The color scales of the neutron intensity are shown on the right side. (b) The evolution of the integrated intensity around the magnetic Bragg peaks derived from (a). The temperature is scaled by $T_N$ = 7 K. The integration intervals for $\theta_i$ (i = 1, ..., 6) are (9.96, 11.56), (13, 14.8), (27.31, 28.5), (28.81, 30.00), (34.13, 35.42), and (35.76, 36.7), respectively. A linear background was assumed in integration.

In order to elucidate the magnetic structure, powder neutron diffraction was performed below 20 K on the D20

instrument at ILL. Whereas single crystal neutron diffraction would be ideal for pinning down the magnetic structure, the required sample volume is difficult to obtain. Nevertheless, as we describe hereafter, powder diffraction still provides essential information to understand this relatively new compound. In **Figure 2a**, we plot the temperature dependence of the low-angle magnetic scattering of CeSiI obtained by subtracting a high-temperature intensity from each $2\theta$. Six magnetic peaks, which are labelled by $\theta_1 \sim \theta_6$ in **Fig. 2a**, are visible below 40° (see Supplementary Fig. 1 for the nuclear refinement). As shown in **Figure 2b**, the positions of these peaks are temperature-independent, while their intensity gradually increases below $T_N$, which is consistent with the presence of a second order phase transition. The quality of the data did not allow for a determination of the critical exponent, however. The increase in the intensity of $\theta_5$ above $T_N$ is affected by a ferromagnetic impurity $CeSi_{1.7}$ with $T_c \sim 11K^{38}$. We note that the change of background below $T_N$ is due to suppression of diffuse scattering by paramagnetic $Ce^{3+}$. We employed a long scan of the 1.6K data subtracted by that of the 7.7 K data for the magnetic structure analysis in order to disentangle the effect of diffuse scattering and magnetic Bragg peak.

| IR | BV | $m_x$ | $m_y$ | $m_z$ |
|---|---|---|---|---|
| $\Gamma_1$ | $\psi_1$ | 0 | 1 | 0 |
| $\Gamma_2$ | $\psi_2$ | 1 | 0 | 0 |
| | $\psi_3$ | 0 | 0 | 1 |

**Table 1** | Irreducible representations (IRs) and basis vectors (BVs) for the space group $P–3m1$ with $k$ = (0.28, 0, 0.19). The decomposition of the magnetic structure representation for the 2c site (0, 0, z) is $\Gamma_{mag} = \Gamma_1 + 2\Gamma_2$. The BV components of one orbit along $a^*$, $b$, and $c$ are shown by $m_x$, $m_y$, and $m_z$, respectively. The other orbit (0, 0, -z) has same BVs.

| Model | IR | $m_{x,1}$ | $m_{y,1}$ | $m_{z,1}$ | $m_{x,2}$ | $m_{y,2}$ | $m_{z,2}$ | $R_p$ (%) | $R_{wp}$ (%) | $\chi^2$ |
|---|---|---|---|---|---|---|---|---|---|---|
| SDW$_b$ | $\Gamma_1$ | 0 | 1.00(1) | 0 | 0 | 1.00(1) | 0 | 39.2 | 28.8 | 6.15 |
| SDW$_{a^*c}$ | $\Gamma_2$ | 0.14(1) | 0 | 1.08(1) | 0.14(1) | 0 | 1.08(1) | 26.9 | 18.2 | 2.59 |
| Cycloid$_1$ | $\Gamma_2$ | 0.07(2)$i$ | 0 | 1.11(1) | 0.07(2)$i$ | 0 | –1.11(1) | 26.7 | 18.6 | 2.76 |
| Cycloid$_2$ | $\Gamma_2$ | 0.71(2)$i$ | 0 | 1.024(9) | 0.71(2)$i$ | 0 | 1.024(9) | 21.8 | 15.9 | 2.26 |

**Table 2** | Results of Rietveld refinement of neutron diffraction pattern. The magnetic structure models are described in the main text. The $m_{i,j}$ represents the component of $j$ site along direction ($i = x, y, z$ and $j = 1, 2$). $R_p$, $R_{wp}$, and $\chi^2$ represent profile factor, weighted profile factor, and reduced chi square, respectively.

Indexing the magnetic peaks produced a unique solution with an incommensurate wavevector of $k \sim$ (0.28, 0, 0.19). In the space group $P–3m1$, the identity and mirror perpendicular to $b$ axis render the propagation vector invariant, and these make up the little group $G_k$. The magnetic representation of a crystallographic site at $2c$ site (0, 0, z) can be decomposed into the irreducible representation $\Gamma_1 + 2\Gamma_2$, for which the projected basis vectors are listed in **Table 1**. While the Ce site is separated into two orbits (0, 0, z) and (0, 0, –z) in $G_k$, the symmetry operation that maps $k$ to $-k$ in the full group relates these two; the real and imaginary coefficients of the basis vectors have opposite and same values between the two orbits, respectively. Inversion and two-fold rotation about the $b$ axis are added to $G_k$ in the full group that also includes time reversal (Supplementary Note 2).

To determine the magnetic structure below $T_N$, we refined the neutron diffraction pattern using all possible magnetic structures described by a single irreducible representation: SDW along the $b$ axis (SDW$_b$), which belongs to $\Gamma_1$ (in BasIreps notation), and SDW (SDW$_{a^*c}$) and counter-rotating cycloid (Cycloid$_1$) in the $a^*c$ plane, which belong to $\Gamma_2$. Cycloid$_1$ has opposite chiralities between the two orbits to maintain the inversion symmetry. We also consider a structure that breaks sublattice symmetry but is an element of the little group and is more common than the counter-rotating cycloid; a co-rotating cycloid (Cycloid$_2$) in the $a^*c$ plane. In all cases, impurity phase $CeSi_{1.7}$ is included in all the refinement and we confirmed that the existence of the secondary phase does not have any impact on the main conclusions of this study. Further details of the magnetic structures and fits are described in Supplementary Note 3. We plot the results of the Rietveld refinement in **Fig.3a-d** and **Table 2**. As shown in **Fig. 3a**, SDW$_b$ is clearly excluded from the candidate magnetic structure because of the poor agreement around the (010)$_-$, (011)$_-$, (011)$_+$, and (01-1)$_+$ reflections. By contrast, SDW$_{a^*c}$,

Cycloid$_1$, and Cycloid$_2$ reproduce the overall behavior of the observed diffraction pattern as indicated by **Figs. 3b-d**. Thus, based on diffraction experiments, we confirm that a common feature of these three magnetic structures, all of which belong to $\Gamma_2$, are a dominant out-of-plane component of the order of 1.0-1.1$\mu_B$. Two of the magnetic structures that yielded the smallest $R$ factors are schematically shown in **Fig. 4a** and **b** as representatives of possible magnetic structures with and without relatively large in-plane magnetic moment.

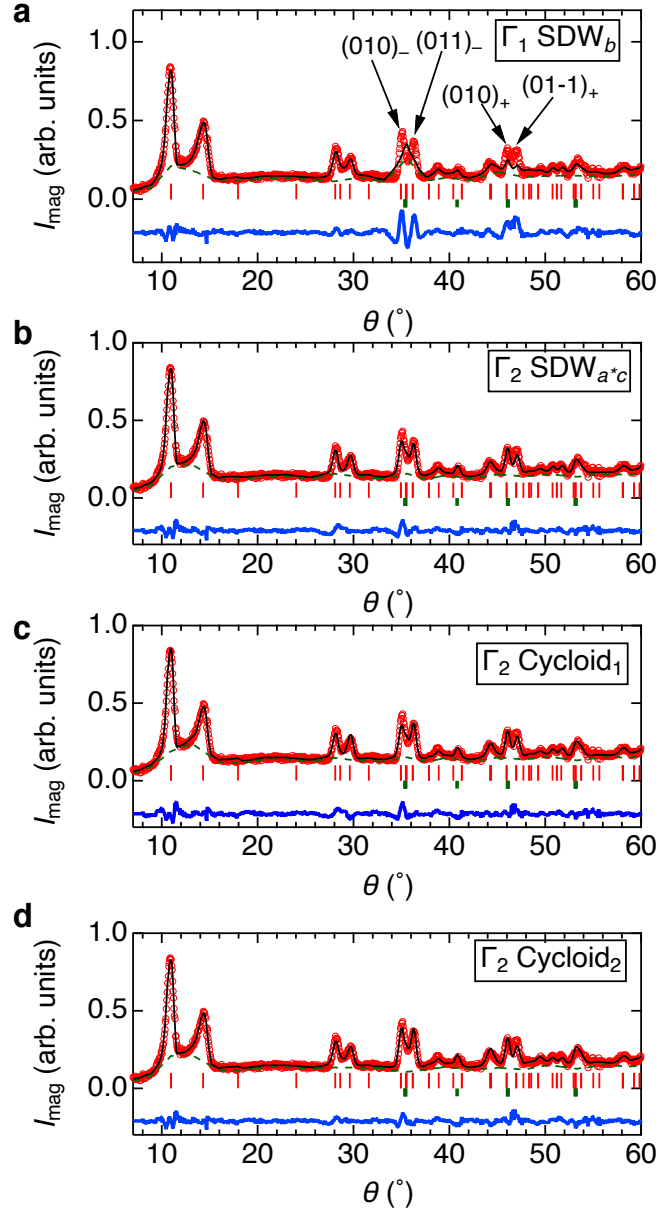

**Figure 3** | Powder neutron diffraction pattern of CeSiI and a Rietveld fitting to the magnetic spin structures $\Gamma_1$ SDW$_b$ (a), $\Gamma_2$ SDW$_{a*c}$ (b), $\Gamma_2$ Cycloid$_1$ (c), and $\Gamma_2$ Cycloid$_1$ (d). The red circles represent data taken at 1.6 K after the subtraction of the 7.7 K data as a reference of nuclear contributions. The black, blue, and green line represent a Rietveld fit, a residual of fitting, and background, respectively. The red thin and thick green bar represents position of magnetic Bragg peaks of CeSiI and CeSi$_{1.7}$, respectively. The arrows indicate the indices of magnetic Bragg peaks, where ($hkl$)$_\pm$ represents a Miller index of ($hkl$)$\pm$(0.28, 0, 0.19) reflection.

Finally, we point out that magnetization process is consistent with the presence of magnetic frustration. **Figure 4c** shows magnetization process along out-of-plane (//$c$) and in-plane (//$ab$) direction. Above 4 T, the magnetization along the $c$ axis reaches ~1.1$\mu_B$, which is close to the expected saturated magnetization of SDW$_{a*c}$ and Cycloid$_2$ model and suggests localized nature of the Ce 4$f$ moment. The magnetization along the $a$ axis does not saturate below 7 T because this is the hard axis. The key feature is seen at 2 K for out-of-plane

(//$c$) case. Between magnetic fields of 2 T and 4 T, two metamagnetic transitions are clearly observed. These anomalies are visible only below $T_N$ and are therefore thought to be associated with the magnetic order. It is important to note that the anomalous behavior should be distinct from a simple spin flop transition in a collinear antiferromagnet, which is usually a single-step process. A similar magnetization process with multiple steps is seen in layered triangular lattice antiferromagnets in insulating systems[39-43], which include $MnI_2$, $NiI_2$, and $CoI_2$. In these compounds, magnetic frustration is supposed to occur due to competing nearest neighbor and further neighbor magnetic coupling, leading to a helical order with in-plane periodicity of ($a$, 0). The magnetic structure in these insulating iodides is analogous to Cycloid$_2$ case (see **table 2**) and competing further neighbor couplings between localized spins are naturally present in metallic systems. As far as we recognize, the observation of multiple magnetization process in a vdW metal is the first report in CeSiI.

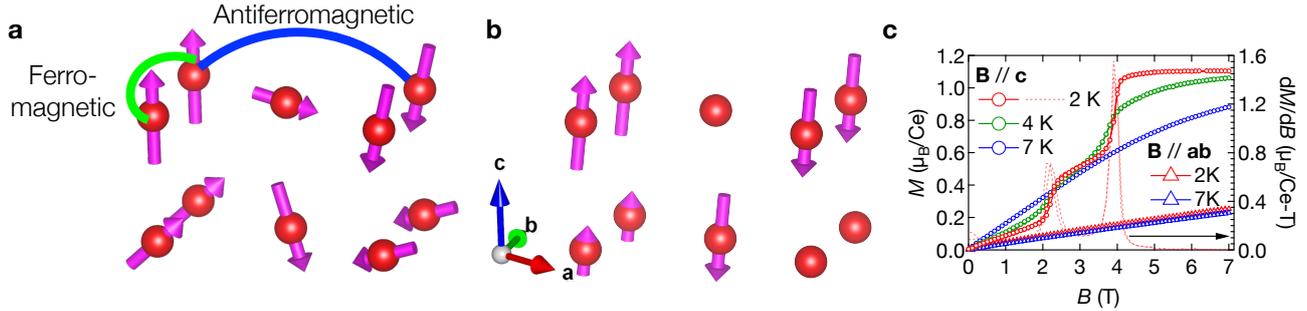

**Figure 4 | Magnetic frustration in CeSiI**
(a,b) Magnetic structures are schematically drawn for the case of Cycloid$_2$ and SDW$_{a*c}$. Red spheres and magenta arrows indicate the magnetic Ce ions and their magnetic moments, respectively. Red, green, and blue arrows indicate the $a$, $b$, and $c$ axes, respectively. (c) Magnetization process of CeSiI. The red, green, and blue circles and red and blue triangles represent magnetization along $c$ axis at 2, 4, and 7 K and along $ab$ plane at 2 and 7 K. respectively. The red dotted lines are derivative of magnetization at 2K, where the peaks indicate the two-step metamagnetic transition.

Our experimental findings collectively point towards CeSiI being a novel material platform hosting magnetic frustration coexisting with itinerant electrons within a vdW material family. In the case of the out-of-plane SDW (**Fig. 4b**), the realization of a quantum disordered state by frustration and reduced dimensionality will be an interesting question to be explored with the help of the exfoliation technique. Further exotic phases can be expected in the case of the co-rotating cycloid case (**Fig. 4a**). For example, the point group of this structure is $m1'$, which allows a finite electric polarization perpendicular to $b$ axis. Even if polarization is allowed by symmetry, the conduction electrons are expected to screen polarization in metallic compounds. However, in the van der Waals metal $WTe_2$, which features a polar-nonpolar structural transition, an out-of-plane electric field can switch the displacement of W when atomically-thin flakes are used[44].

In summary, we investigated CeSiI using resistivity, magnetometry measurements and neutron diffraction experiments on both powder and single crystal samples. Through these, we demonstrated that CeSiI is a novel material platform to search exotic phenomena arisen from rich interplay between spin, charge and lattice degree of freedom, for example, by using exfoliation-based device fabrication techniques.

**Acknowledgement**
We thank Takeshi Yajima, and Hajime Ishikawa for advice on the synthesis of CeSiI, Dmitry Khalyavin for help with the determination of the magnetic structure, and Rieko Ishii, Daichi Ueta, and Wonjong Lee for helping the experiments. This work was carried out by the joint research in the Institute for Solid State Physics, the University of Tokyo.

**Method**
**Single crystal growth**
Single crystals of CeSiI were synthesized by a high-temperature-solid-state reaction following ref. 33. The handling of the elements and compounds was performed in an argon-filled glovebox. $CeI_3$ was made by reaction of stoichiometric amounts of each element: Ce and $I_2$ were sealed in an evacuated quartz tube in vacuo and reacted at 300˚C for 24 hours followed by subsequent heating at 900˚C for 24 hours to complete the reaction. The quartz tube was then placed vertically with one end outside the furnace and heated again at 900˚C in order to facilitate the sublimation of polycrystalline $CeI_3$. Yellow needle-like crystals of $CeI_3$ were formed at the colder end of the quartz tube. Stoichiometric amounts of $CeI_3$, Ce, and Si in a total weight of 1g were sealed in a Nb tube with a diameter of 1 cm and length of 15 cm by arc welding. The sealed Nb tube was sealed in a quartz tube to prevent oxidation of Nb at elevated temperatures. The sealed quartz tube was then placed horizontally in a two-zone furnace. The cold and hot ends were kept 800˚C and 1000˚C, respectively. After a week, shiny bronze plate-like crystals with a maximum dimension of 2 x 2 x 0.05 mm$^3$ formed at the hot end. The synthesis of powder CeSiI was performed at 1000˚C for four days.

**X-ray diffraction, magnetometry, heat capacity, and resistivity measurement**
Single crystals of CeSiI could be indexed by a trigonal space group $P$-3$m$1 with lattice constants of $a$ = 4.1872(1) Å and $c$ = 11.7036(4) Å at 300 K using a single crystal X-ray diffractometer Bruker Venture D8 (Mo K$\alpha$, $\lambda$ = 0.71069 Å). DC SQUID Magnetometry measurements were performed using MPMS-3 (Quantum Design). A single crystal was placed on a diamagnetic sapphire substrate and coated with paraffin. Resistivity was measured by conventional four-probe method using the resistivity option of PPMS (Quantum Design). An as-grown polycrystalline chunk of CeSiI was attached to indium electrodes and gold wires.

**Powder Neutron Diffraction experiment and data analysis.**
Powder neutron diffraction of was performed on the high-flux D20 thermal diffractometer at Institut Laue-Langevin at temperatures from 20 K to 1.6 K using a constant wavelength $\lambda_1 \sim 2.4$ Å[45]. The powder sample was sealed in a vanadium can with an indium seal to prevent oxidation. The magnetic peaks were indexed by the *k*-search software in the FULLPROF suite[46]. The analysis of the irreducible representations is performed using SARAh suite software (little group)[47] and ISODISTORT software (full group)[48,49]. Structural and magnetic refinement of powder diffraction data was performed by Fullprof[46].

**Data availability**
The data files are also available from the corresponding author upon request.

# Supporting information for "Magnetic frustration in a van der Waals metal CeSiI"

Ryutaro Okuma,[1,4] Clemens Ritter,[2] Gøran J. Nilsen,[3] Yoshinori Okada[1]
[1]Okinawa Institute of Science and Technology Graduate University, Onna-son, Okinawa, 904-0495, Japan
[2]Institut Laue-Langevin, 71 avenue des Martyrs, 38042 Grenoble, France
[3]ISIS Neutron and Muon Source, Science and Technology Facilities Council, Didcot OX11 0QX, United Kingdom
[4]Present address: Clarendon Laboratory, University of Oxford, Oxford, OX1 3PU, UK

**Supplementary Note 1    Nuclear structure refinement of powder neutron diffraction data**
We show here that neutron analysis for drawing main conclusion is not altered due to impurity phases in our samples. **Figure S1** shows the result of nuclear Rietveld refinement in Fullprof[1] using data at 19.4 K considering four phases according to reported structural data. It successfully converged to $R_{wp}$ = 8.76% and $R_p$ = 7.45%. The refined value of molar ratio of the sample was CeSiI : CeI$_3$ : CeSi$_{1.7}$ : CeOI = 56.4(6) : 25.3(4) : 14.4(3) : 3.9(1). As CeI$_3$ and CeOI are paramagnetic above 2K and CeSi$_{1.7}$ is ferromagnetic below ~11 K[2], magnetic scattering away from nuclear scattering should originate purely from magnetic moments on CeSiI.

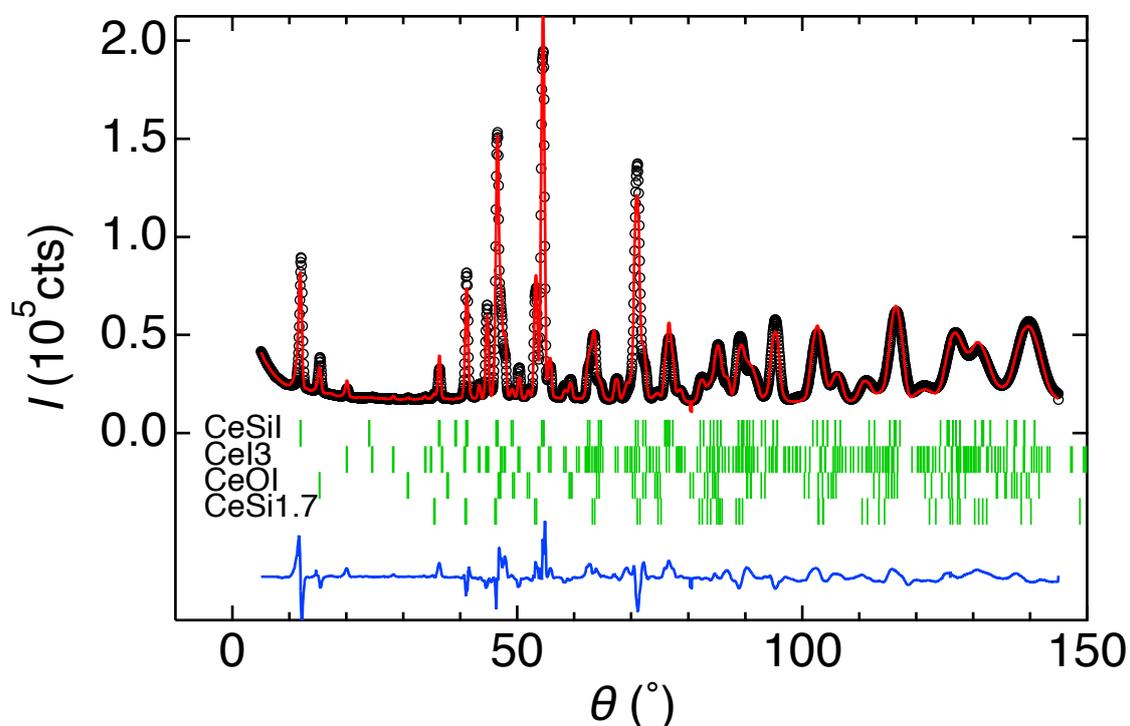

**Supplementary Fig. 1 | Powder neutron profile of CeSiI at 19.4 K fitted by the Rietveld method, showing observed (black circle), calculated (red line), and difference (blue line).** The green ticks represent the positions of the Bragg reflections of CeSiI and impurity phases: CeI$_3$, CeOI, and CeSi$_{1.7}$.

| Fomula | | | | CeSiI | |
|---|---|---|---|---|---|
| Space group | | | | $P$-3$m$1 | |
| $a$ / Å | | | | 4.1697(1) | |
| $c$ / Å | | | | 11.6581(4) | |
| $V$ / Å$^3$ | | | | 175.537(9) | |
| $Z$ | | | | 2 | |
| Atom | Wyck. | $x$ | $y$ | $z$ | Occ. |
| Ce1 | 2$c$ | 0 | 0 | 0.1751(6) | 1 |
| I1 | 2$d$ | 1/3 | 2/3 | 0.3548(4) | 1 |
| Si1 | 2$d$ | 1/3 | 2/3 | 0.9924(5) | 1 |

| Fomula | | | | CeI$_3$ | |
|---|---|---|---|---|---|
| Space group | | | | $Cmcm$ | |
| $a$ / Å | | | | 4.3636(3) | |
| $b$ / Å | | | | 13.917(1) | |
| $c$ / Å | | | | 9.9333(7) | |
| $V$ / Å$^3$ | | | | 603.23(8) | |
| $Z$ | | | | 4 | |
| Atom | Wyck. | $x$ | $y$ | $z$ | Occ. |
| Ce1 | 4$c$ | 0 | 3/4 | 1/4 | 1 |
| I1 | 4$c$ | 0 | 0.0803(7) | 1/4 | 1 |
| I2 | 8$f$ | 0 | 0.3541(5) | 0.0700(7) | 1 |

| Fomula | CeSi$_{1.7}$ |       |     |          |      |
|:------:|:------------:|:-----:|:---:|:--------:|:----:|
| Space group | *Imma* |    |     |          |      |
| *a* / Å | 4.1195(4) |     |     |          |      |
| *b* / Å | 4.1835(4) |     |     |          |      |
| *c* / Å | 13.883(1) |     |     |          |      |
| *V* / Å$^3$ | 239.26(4) |  |   |          |      |
| *Z* | 4 |              |     |          |      |
| Atom | Wyck. | *x* | *y* | *z* | Occ. |
| Ce1 | 4*e* | 0 | 1/4 | 0.124(1) | 1 |
| Si1 | 4*e* | 0 | 1/4 | 0.533(1) | 0.85 |
| Si2 | 4*e* | 0 | 1/4 | 0.702(1) | 0.85 |

| Fomula | CeOI |       |     |       |      |
|:------:|:----:|:-----:|:---:|:-----:|:----:|
| Space group | *P4/nmm* | | | | |
| *a* / Å | 4.1051(5) | | | | |
| *c* / Å | 9.100(2) | | | | |
| *V* / Å$^3$ | 153.35(5) | | | | |
| *Z* | 2 | | | | |
| Atom | Wyck. | *x* | *y* | *z* | Occ. |
| Ce1 | 2*c* | 1/4 | 1/4 | 0.131 | 1 |
| O1 | 2*a* | 3/4 | 1/4 | 0 | 1 |
| I1 | 2*c* | 1/4 | 1/4 | 0.672 | 1 |

**Supplementary Table 1 | Structural parameters of CeSiI, CeSi$_{1.7}$, CeI$_3$, CeOI, and obtained from the Rietveld refinements.** Isotropic displacement parameter $B_{iso}$ is fixed to 0.5/Å$^2$ for all atoms.

**Supplementary Note 2  Symmetry analysis of the magnetic structure**

Details of the symmetry analysis of the magnetic structure is given in this section. CeSiI crystallizes in a trigonal space group *P-3m1* with lattice constants *a* = 4.1697(1) Å and *c* = 11.6581(4) Å at 20 K and Ce ions occupy single Wyckoff position 2*c* as described in the Supplementary Note 1. To comply with the symmetry of the single-*k* magnetic structure described by *k* = 0.28*a** + 0.2*c**, we use (*a** *b c*) as principal axes in the following discussion. The magnetic moment at the position *r* = (*x*, *y*, *z*) can be expressed as

$$S(r) = e^{ik \cdot r}(m_x, m_y, m_z) + e^{-ik \cdot r}(m_x^*, m_y^*, m_z^*).$$

Here $m_x$, $m_y$, $m_z$, and * correspond to the component along *a**, *b*, and *c* and a complex conjugate operator, respectively. Symmetry of the magnetic structure can be characterized by the symmetry operations in *P-3m11'* that transform *k*•*r* to ±*k*•*r*, namely,

$$1': (x, y, z, m_x, m_y, m_z) \rightarrow (x, y, z, -m_x, -m_y, -m_z),$$
$$m_b: (x, y, z, m_x, m_y, m_z) \rightarrow (x, -y, z, -m_x, m_y, -m_z),$$
$$2_b: (x, y, z, m_x, m_y, m_z) \rightarrow (-x, y, -z, -m_x^*, m_y^*, -m_z^*),$$

where 1', $m_b$, and $2_b$ describe time-reversal operation, mirror operation perpendicular to the *b* axis, and two-fold rotation about the *b* axis, respectively. Superspace groups of the candidate magnetic structures are obtained by ISODISTORT[3,4] as shown in Supplementary Table 2 and Fig. 2.

| IR | Superspace group | Magnetic point group | $S_+$ | $S_-$ | Types of magnetic structures |
|:--:|:----------------:|:--------------------:|:-----:|:-----:|:----------------------------:|
| $\Gamma_1$ | *C2/m*1'(α, 0, β)00*s* | 2/*m*1' | (0, *m*, 0) | (0, *m**, 0) | SDW // *b* axis |
|            | *Cm*1'(α, 0, β)0*s* | *m*1' | (0, *m*, 0) | (0, *u*, 0) | SDW // *b* axis |
| $\Gamma_2$ | *C2/m*1'(α, 0, β)0*ss* | 2/*m*1' | (*m*, 0, *u*) | −(*m**, 0, *u**) | SDW and counter-rotating cycloid // *a*\**c* |
|            | *Cm*1'(α, 0, β)*ss* | *m*1' | (*m*, 0, *u*) | (*v*, 0, *w*) | SDW and co-rotating cycloid // *a*\**c* |

**Supplementary Table 2 | Superspace groups of single-*k* magnetic structures with the incommensurate propagation vector *k* = 0.28*a** + 0.2*c**.** $S_\pm$ denotes the allowed ($m_x$, $m_y$, $m_z$) at (0, 0, ±*z*) in the given superspace group and *m*, *u*, *v*, and *w* are complex numbers. The *s* in superspace groups denotes the shift of the internal coordinate by a factor of π.

In *C2/m*1'(α, 0, β)00*s* and *C2/m*1'(α, 0, β)0*ss*, two-fold rotation related the two orbits inside a unit cell and inversion symmetry is retained whereas in *Cm*1'(α, 0, β)0*s* and *Cm*1'(α, 0, β)*ss*, two orbits are not related by symmetry and only mirror symmetry is present as shown in Supplementary Fig. 2. In terms of types of magnetic structure, *C2/m*1'(α, 0, β)00*s* allows only a collinear spin density wave order parallel to *b* axis and both orbits have same amplitude. *Cm*1'(α, 0, β)0*s* also allows only a collinear SDW but the amplitude between the sites can be different. *C2/m*1'(α, 0, β)0*ss* in general describes elliptic counter rotating cycloids in the sense that the chirality defined as *b*•$S_r$×$S_{r+a}$ takes the opposite sign and same amplitude between the two orbits, which includes SDW parallel to the *b* axis. *Cm*1'(α, 0, β)*ss* can describe more general elliptic cycloids including co-rotating cycloids.

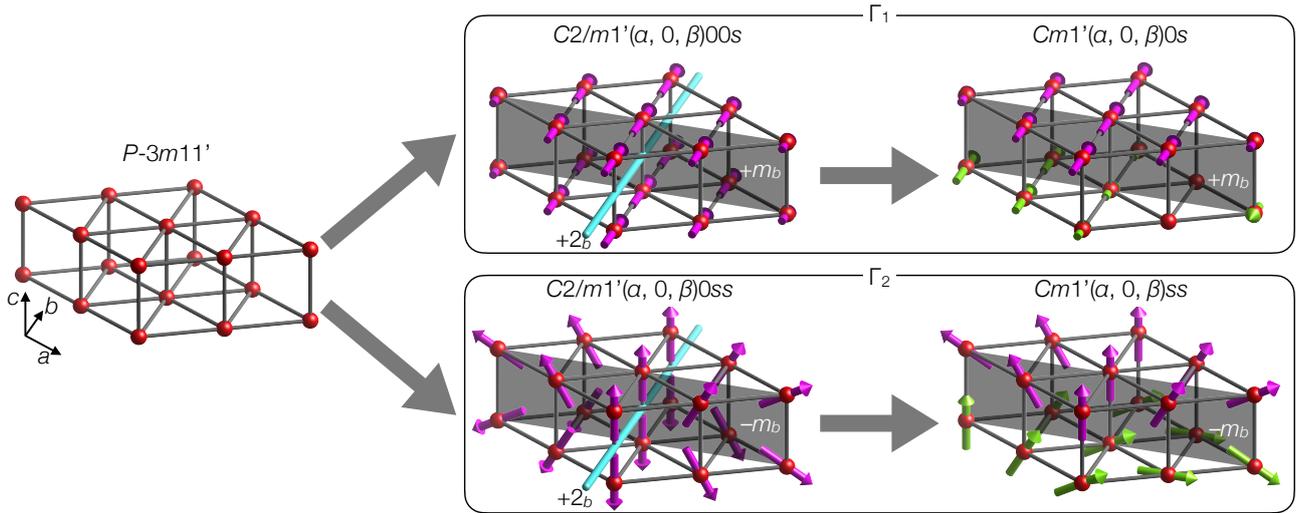

Supplementary Fig. 3 | Typical magnetic structure allowed in the superspace group given in Supplementary Table 2. The red sphere and purple and green arrows represent Ce ions and magnetic moments with different sites, respectively. Sky-blue bond and shaded plane indicate two-fold rotation axis and mirror plane, respectively. The sign indicates phase shift of the internal coordinate; it is multiplied to the magnetic moment after of the symmetry operation is performed to keep the magnetic structure unchanged. SDW, counter-rotating cycloid, and co-rotating cycloids are sketched as examples for $C2/m1'(\alpha, 0, \beta)00s$ and $Cm1'(\alpha, 0, \beta)0s$, $C2/m1'(\alpha, 0, \beta)00s$, and $Cm1'(\alpha, 0, \beta)ss$, respectively.

**Supplementary Note 3    Magnetic structure refinement**

To construct physical models of the magnetic structure in CeSiI, we considered typical magnetic structures allowed in the superspace groups of Supplementary Table 2, which is discussed in the main text. Each parameter used to describe the magnetic structures is shown in Supplementary Table3.

| Model | Superspace group | $S_+$ | $S_-$ |
|---|---|---|---|
| $SDW_b$ | $C2/m1'(\alpha, 0, \beta)00s$ | $e^{-i\varphi/2} (0, M_y, 0)$ | $e^{i\varphi/2} (0, M_y, 0)$ |
| $SDW_{a*c}$ | $C2/m1'(\alpha, 0, \beta)0ss$ | $e^{-i\varphi/2} (M_x, 0, M_z)$ | $e^{i\varphi/2} (M_x, 0, M_z)$ |
| $Cycloid_1$ | $C2/m1'(\alpha, 0, \beta)0ss$ | $e^{-i\varphi/2} (iM_x, 0, M_z)$ | $e^{i\varphi/2} (iM_x, 0, -M_z)$ |
| $Cycloid_2$ | $Cm1'(\alpha, 0, \beta)ss$ | $e^{-i\varphi/2} (iM_x, 0, M_z)$ | $e^{i\varphi/2} (iM_x, 0, M_z)$ |

**Supplementary Table 3 | Magnetic structure models used in the Rietveld refinement.** $S_\pm$ denotes the allowed ($m_x$, $m_y$, $m_z$) at $(0, 0, \pm z)$ in the given superspace group and $M_x$, $M_y$, $M_z$, $\varphi$, and $\theta$ are real numbers.

We note that there are three features that require careful treatment in the magnetic Rietveld refinement of the powder diffraction data while they do not affect the main conclusion that the magnetic structure has an incommensurate dominant easy-axis component: ferromagnetic impurity phase $CeSi_{1.7}$, asymmetric peak broadening and diffuse scattering at low angles. First, the ferromagnetic reflections of $CeSi_{1.7}$ were included in all the analysis assuming that the Ce moment points in the <100> direction[1]. Second, asymmetric peak broadening, which manifests itself at $(000)_\pm$ and $(001)_-$ peaks, indicates finite correlation length along the out-of-plane direction due to two-dimensional character of the magnetism. This effect is simulated by plate-like magnetic domains via Size-model function in Fullprof[1]. Finally, presence of diffuse scattering is inferred from the residual intensity around 15°, between $(000)_\pm$ and $(001)_-$ reflections, which cannot totally be reproduced by plate-like domains. For the background, we initially used a linear interpolation of fixed points in the regions without magnetic Bragg peaks. Then to reproduce the actual background containing broad peaks of diffuse scattering, we also refined the background intensity of intermediate positions, which include magnetic Bragg peaks. The results of the refinement with the fixed background points and the refined background points are presented in Supplementary Tables 4 and Supplementary Fig. 4 and Supplementary Table 5 and Supplementary Fig.5, respectively.

In the fitting without refinable background, the obtained $M_z$ exceeds the saturation field of $1.1\mu_B$ and $\chi^2$ is much larger than 1 due to poor agreement near $(000)_\pm$ and $(001)_-$ peak. By contrast, in the fitting with refinable background, $M_z$ is closed to the observed value in $\Gamma_2$ models and $\chi^2$ is greatly improved. The refined $CeSi_{1.7}$ moment also agrees well with the reported value of $0.5\mu_B{}^2$ regardless of the background refinement. Therefore, fittings with the refined background are more likely to capture the feature of the magnetic structure and $Cycloid_2$ is the primary candidate of the magnetic structure. However, small difference in the agreement factor between the collinear and noncollinear orders certainly calls for polarized neutron diffraction experiments using single crystals in order to measure the chiral term $iM_q \times M_q^*$, which is proportional to spin chirality. Although the limitation of the crystal size hinders such a measurement at the current stage, a large single crystal will uncover the magnetic structures even under magnetic fields and thus shed light on the origin of the frustration in CeSiI.

| Model | $M_x$ or $M_y$ | $M_z$ | $\varphi$ | $\xi$ | $k_{a^*}$ | $k_{c^*}$ | $M_{CeSi_{1.7}}$ | $R_p$ | $R_{wp}$ | $R_{exp}$ | $\chi^2$ |
|---|---|---|---|---|---|---|---|---|---|---|---|
| SDW$_b$ | 1.157(7) | − | 0.351(3) | 15.7(5) | 0.2784(1) | 0.2029(8) | 0.65(2) | 36.5 | 31.7 | 9.18 | 11.9 |
| SDW$_{a^*c}$ | 0.25(1) | 1.262(6) | 0.363(2) | 14.0(3) | 0.27807(5) | 0.2032(4) | 0.43(1) | 24.7 | 18.4 | 9.17 | 4.0 |
| Cycloid$_1$ | 0.22(3) | 1.288(7) | 0.187(1) | 13.6(3) | 0.27824(6) | 0.2059(5) | 0.46(1) | 26.6 | 20.8 | 9.17 | 5.1 |
| Cycloid$_2$ | 0.40(4) | 1.27(1) | 0.179(2) | 12.5(3) | 0.27800(6) | 0.2047(5) | 0.42(2) | 27.9 | 21.6 | 9.17 | 5.5 |

**Supplementary Table 4 | Magnetic structure refinement with fixed background.** $M$, $\theta$, and $\varphi$ are parameters that define the magnetic structure in Supplementary Table3. $\xi$ is anisotropic Lorentzian contribution of particle size for a plate-like coherent domain[3]. $k_{a^*}$, $k_{c^*}$ are $a^*$ and $c^*$ component of the propagation vector. $M_{CeSi_{1.7}}$ is length of ferromagnetic moment of CeSi$_{1.7}$. $R_p$, $R_{wp}$, and $R_{exp}$, $\chi^2$ represent profile factor, weighted profile factor, expected weighted profile factor, and reduced chi square, respectively.

| Model | $M_x$ or $M_y$ | $M_z$ | $\varphi$ | $\xi$ | $k_{a^*}$ | $k_{c^*}$ | $M_{CeSi_{1.7}}$ | $R_p$ | $R_{wp}$ | $R_{exp}$ | $\chi^2$ |
|---|---|---|---|---|---|---|---|---|---|---|---|
| SDW$_b$ | 1.00(1) | − | 0.344(5) | 10.4(6) | 0.27833(8) | 0.1936(6) | 0.37(3) | 39.2 | 28.8 | 11.6 | 6.2 |
| SDW$_{a^*c}$ | 0.14(1) | 1.08(1) | 0.355(3) | 7.5(3) | 0.27809(5) | 0.1935(4) | 0.52(1) | 26.9 | 18.2 | 11.3 | 2.6 |
| Cycloid$_1$ | 0.07(1) | 1.11(1) | 0.175(2) | 6.5(3) | 0.27808(5) | 0.1938(4) | 0.53(1) | 26.7 | 18.6 | 11.2 | 2.8 |
| Cycloid$_2$ | 0.71(2) | 1.024(9) | 0.161(2) | 8.1(3) | 0.27806(4) | 0.1936(3) | 0.55(1) | 21.8 | 15.9 | 10.6 | 2.3 |

**Supplementary Table 5 | Magnetic structure refinement with refined background.** The refinable background consists of 36 points between 9 and 99˚ in equal spacing of 2.5˚.

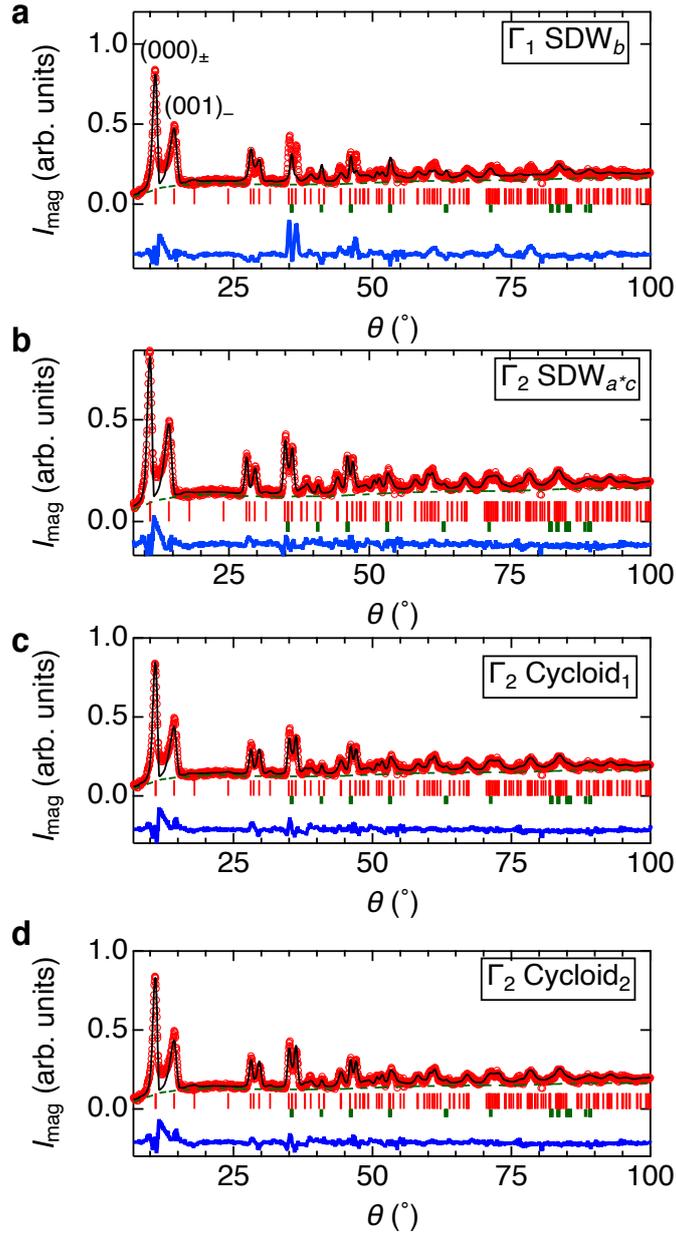

**Supplementary Figure 4 | Powder neutron diffraction pattern of CeSiI and a Rietveld fitting to the magnetic spin structures $\Gamma_1$ SDW$_b$ (a), $\Gamma_2$ SDW$_{a^*c}$ (b), $\Gamma_2$ Cycloid$_1$ (c), and $\Gamma_2$ Cycloid$_1$ (d) with fixed background parameters.** Fitting regions are between 8° and 100°. The red circle represents data taken at 1.6 K after the subtraction of the 7.7 K data as a reference of nuclear contributions. The black, blue, and green line represent a Rietveld fit, a residual of fitting, and background, respectively. The red thin and thick green bar represents position of magnetic Bragg peaks of CeSiI and CeSi$_{1.7}$, respectively. The arrows indicate the indices of magnetic Bragg peaks, where $(hkl)_\pm$ represents a Miller index of $(hkl)\pm(0.28, 0, 0.20)$ reflection.

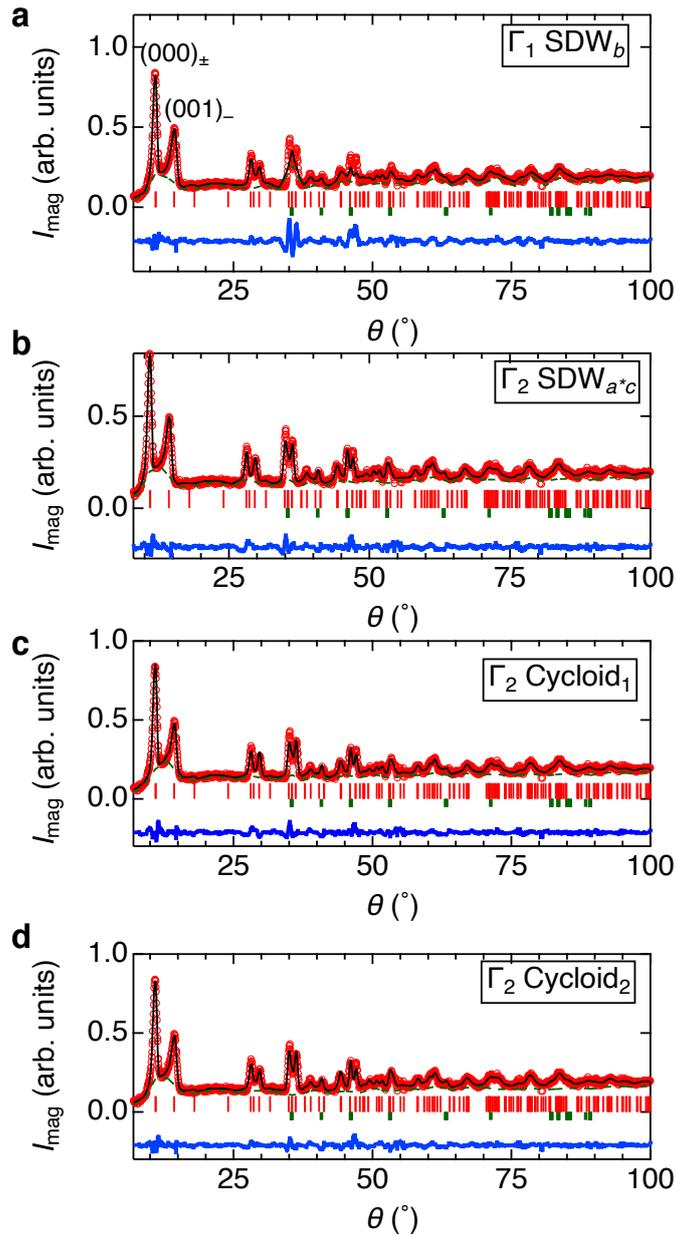

**Supplementary Figure 5 | Powder neutron diffraction pattern of CeSiI and a Rietveld fitting to the magnetic spin structures with refined background parameters.**

**Supplementary References**